\documentclass[final]{svjour3}
\usepackage{rotating}
\usepackage{amssymb}
\usepackage{mathptmx}
\usepackage[english]{babel}
\usepackage{amsmath}
\usepackage{graphicx}
\usepackage[colorinlistoftodos]{todonotes}
\usepackage{hyperref}
\usepackage{amsmath}
\usepackage{setspace}
\usepackage{cite}
\usepackage{subcaption}
\usepackage[font={small}]{caption}
\makeatletter
\journalname{Journal of Low Temperature Physics}

\begin{document}

\setlength{\abovedisplayskip}{3pt}
\setlength{\belowdisplayskip}{3pt}

\newcommand{\hdblarrow}{H\makebox[0.9ex][l]{$\downdownarrows$}-}
\title{Operation of a superconducting nanowire in two detection modes: KID and SPD}

\author{Edward Schroeder$^\dagger$ \and Philip Mauskopf$^{\dagger\ast}$ \and\\ Hamdi Mani$^\ast$ \and Sean Bryan$^\ast$ \and\\ Karl K. Berggren$^\star$ \and Di Zhu$^\star$}

\institute{$^\dagger$Department of Physics, Arizona State University,\\ Tempe, AZ 85281, United States\\
$^\ast$School of Earth and Space Exploration, Arizona State University,\\ Tempe, AZ 85281, United States\\
$^\star$Department of Electrical Engineering and Computer Science, Massachusetts Institute of Technology,\\ Cambridge, MA 02139, United States\\
\email{erschroe@asu.edu}}

\maketitle

\begin{abstract}

We present the performance of a superconducting nanowire that can be operated in two detection modes: i) as a kinetic inductance detector (KID) or ii) as a single-photon detector (SPD). Two superconducting nanowires developed for use as single-photon detectors (SNSPDs) are embedded as the inductive (L) component in resonant inductor/capacitor (LC) circuits coupled to a microwave transmission line. The capacitors are low loss commercial chip capacitors and limit the internal quality factor of the resonators to approximately $Q_i = 170$. The resonator quality factor, $Q_r \simeq 23$, is dominated by the coupling to the feedline and limits the detection bandwidth to on the order of 1MHz. When operated in KID mode, the detectors are AC biased with tones at their resonant frequencies of 45.85 and 91.81MHz. In the low-bias, standard KID mode, a single photon produces a hot spot that does not turn an entire section of the line normal but only increases the kinetic inductance. In the high-bias, critical KID mode, a photon event turns a section of the line normal and the resonance is destroyed until the normal region is dissipated. When operated as an SPD in Geiger mode, the resonators are DC biased through cryogenic bias tees and each photon produces a sharp voltage step followed by a ringdown signal at the resonant frequency of the detector which is converted to a standard pulse with an envelop detector. We show that AC biasing in the critical KID mode is inferior to the sensitivity achieved in DC-biased SPD mode due to the small fraction of time spent near the critical current with an AC bias.


\end{abstract}
\pagebreak
\section{Introduction}

Superconducting nanowire devices are a promising technology for use in astronomy, quantum optics, communications, and the life sciences. One prominent application in astronomy for these devices is intensity interferometry\cite{Schroeder2016}. This application requires quick and accurate detection of photons in the radiation field from stellar objects and is achieved by operating the superconducting nanowire as a single photon detector (SNSPD). SNSPDs are biased close to the switching current such that a single photon carries enough energy to drive a section of the nanowire normal and produce a countable voltage pulse \cite{2012SuScT..25f3001N,Zhu:16,0953-2048-28-11-114003}. 

To operate an array of SNSPDs with this traditional approach, each device needs to be wired individually from cryogenic temperatures to a room temperature environment. This increases system complexity and heat load on the cold-plate, resulting in higher power consumption. Implementing the nanowire in a resonant circuit allows the multiplexing scheme already well-developed for kinetic inductance detectors (KIDs) to be used. Since the capacitive portion of the resonant circuit is easily tuned, the resonant frequency of each pixel can be uniquely read out on the same feedline. Additionally, the resonators can be tuned by varying the dimensions of the nanowire\cite{0957-4484-21-44-445202}. 

It is also possible to bias the nanowires with RF-power instead of DC\cite{7399730}. Doerner et al. demonstrated an early two-pixel proof-of-concept for multiplexed SNSPDs biased with RF-power. Their resonant circuits consisted of a coupling capacitor in addition to the inductive nanowire. More recently, a RF-biased 16-pixel array was developed and tested by this group. The pixels consisted of an coupling capacitor, a parallel resonator formed with a capacitor and an inductor, and the photo-sensitive nanowire\cite{doi:10.1063/1.4993779}. 

A time-tagged readout has been developed by Hofherr et al.\cite{6459543} by coupling SNSPDs in an array with a superconducting delay line. This method involves the use of one bias supply for the whole chain and produces an effective temporal resolution of photon events. Another method for building a nanowire-based imager has been carried out by Zhao et al.\cite{Zhao2017}. Their imager is based on a single superconducting nanowire that is read out at both ends in order to determine the location along the wire and the time at which a single photon event occurred.
%
%

In this paper we consider two modes of SNSPD operation: linear mode and Geiger mode. Linear mode detectors give a response proportional to fluctuations in absorbed power and include calorimeters such as transition edge sensors (TES), and KIDs. Given enough sensitivity and speed, these detectors can resolve individual photons and measure photon energy. Geiger-mode detectors such as photomultipliers or avalanche photodiodes 
are threshold detectors and produce uniform pulses in response to absorption of photons. We operate the nanowire in a resonant LC circuit as a KID (linear mode) with low-AC readout power and high-AC readout power. We also operate the nanowire with a DC current bias embedded in the resonant circuit and convert the ringing signal to a standard, countable pulse (Geiger mode). 

\section{Experimental Apparatus}

\begin{figure*}[t!]
    \centering
    \begin{subfigure}[t]{1\textwidth}
        \centering
        \includegraphics[scale=0.3]{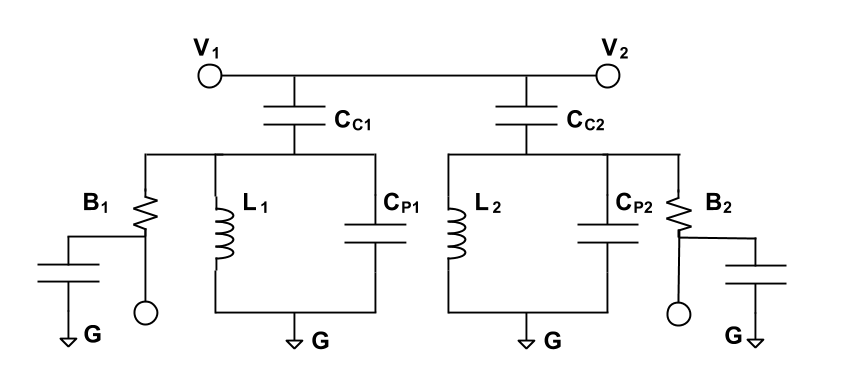}
        \caption{}
        \label{fig:diag1}
    \end{subfigure}
    
    \begin{subfigure}[t]{0.4\textwidth}
        \centering
        \includegraphics[scale=0.05]{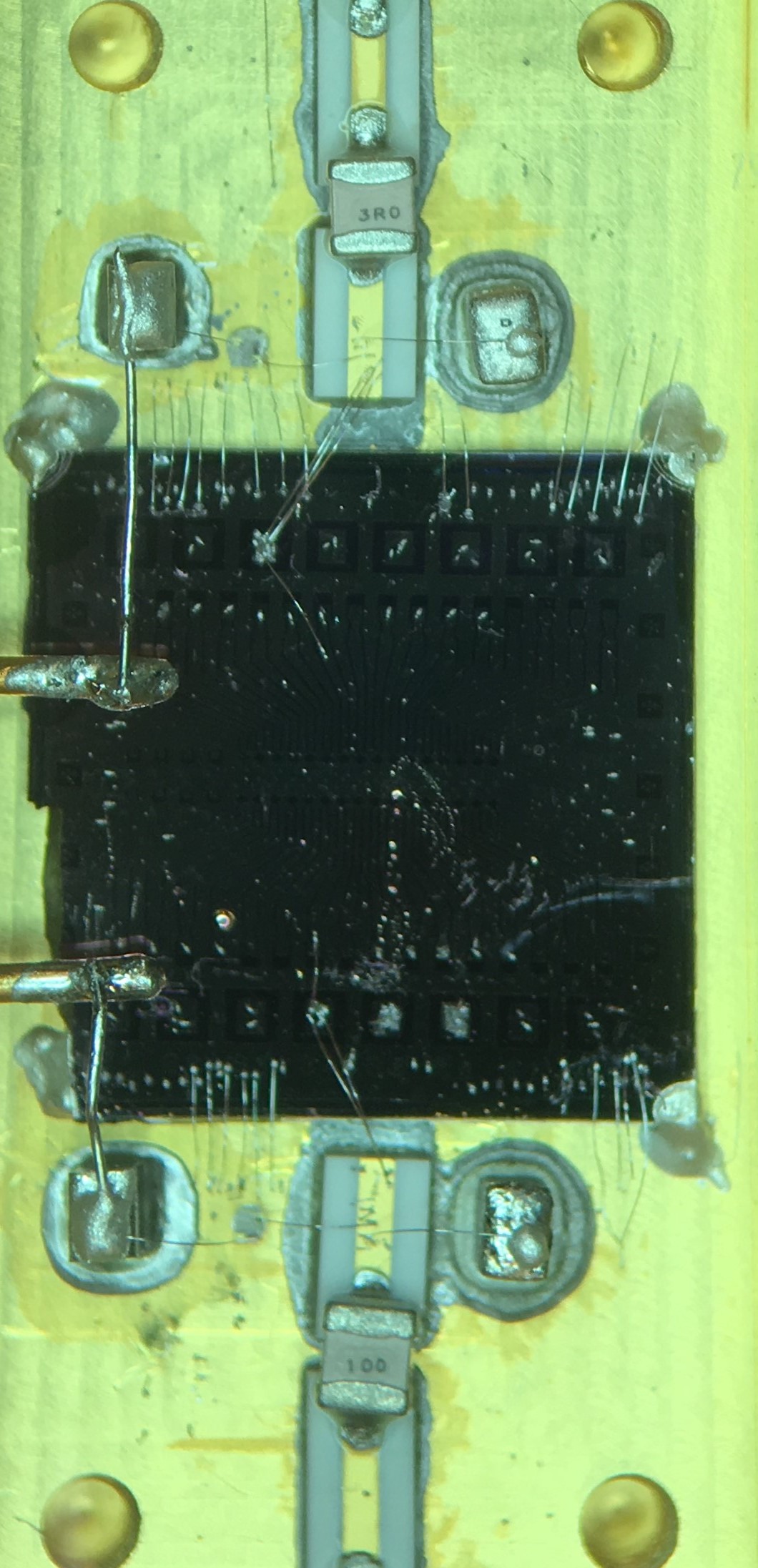}
        \caption{}
        \label{fig:diag2}
    \end{subfigure}    
     ~
    \begin{subfigure}[t]{0.4\textwidth}
        \centering
        \includegraphics[scale=0.09]{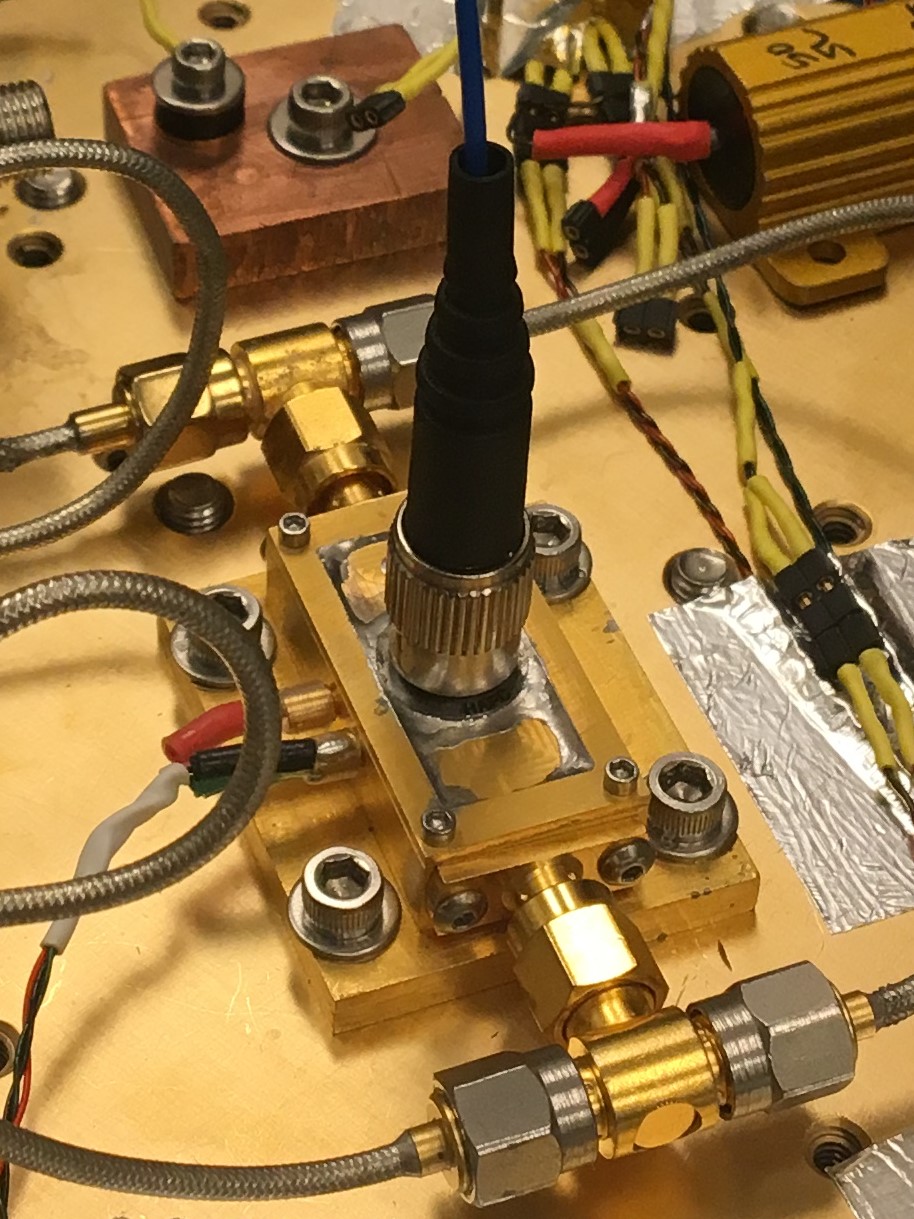}
        \caption{}
        \label{fig:diag3}
    \end{subfigure}
    \caption{(a) Circuit diagram of two devices on the same feedline. 
    (b) Micro-assembly inside the detector package. (c) Fiber-coupled detectors mounted inside the cryostat on the cold-plate.}
    \label{fig:SNR_diag}
\end{figure*}

A Sumitomo RP-082 closed-cycle cryogenic cooler was used to cool the resonators to a bath temperature of $T = 2.7K$. A cryogenically cooled $30dB$ attenuator was used at the RF input of the resonator package and a cryogenically cooled low noise amplifier (LNA) developed at ASU\footnote{Datasheet: http://thz.asu.edu/products.html} was used to boost the signal with $30dB$ of power gain at the output with $4K$ of input referred noise temperature.

A $1.3\mu m$ LED was used to apply optical power and was biased with a Keithley 2400 series precision source measurement unit (SMU). The microwave response of the devices were read out with an Agilent E5072A vector network analyzer (VNA). In Geiger mode, the devices were biased with a temperature stabilized, battery operated, precision, constant current source made at ASU. The waveforms were recorded using a Tektronix TDS 7104 oscilloscope and the pulses were counted using a Tektronix FCA 3100 frequency counter.

Figure \ref{fig:diag1} shows the circuit diagram of the two devices. The physical assembly is shown in Figure \ref{fig:diag2}. The central chip contains the nanowires on a silicon nitride-on-silicon substrate and was fabricated at MIT by the Quantum Nanostructures and Nanofabrication Group. The nanowire consists of serially connected parallel sections with a nominal width of $60nm$. The total length is $500\mu m$ and the thickness is $\sim 4nm$. The inductances of the nanowires are $L_1=731nH$ and $L_2=586nH$. These values are obtained from the measured values of the resonances. 
%
%

Each nanowire is connected to their respective circuits with $25\mu m$ diameter aluminum wire and has two associated surface mount device (SMD), high Q capacitors. For the first device, $C_{C1}=3pF$ and $C_{P1}=1pF$. For the second device, $C_{C2}=10pF$ and $C_{P2}=10pF$. This concept is similar to that proposed by Doerner et al., where instead of SMD capacitors, interdigital capacitors are used. Figure \ref{fig:diag3} shows the mounted package coupled to a multi-mode fiber optic cable.
\vspace{-0.5cm}

\section{Operation in Linear Mode}

\subsection{Standard Kinetic Inductance Detector}

\begin{figure*}[t!]
    \centering
    \begin{subfigure}[t]{0.5\textwidth}
        \centering
        \includegraphics[width=2in]{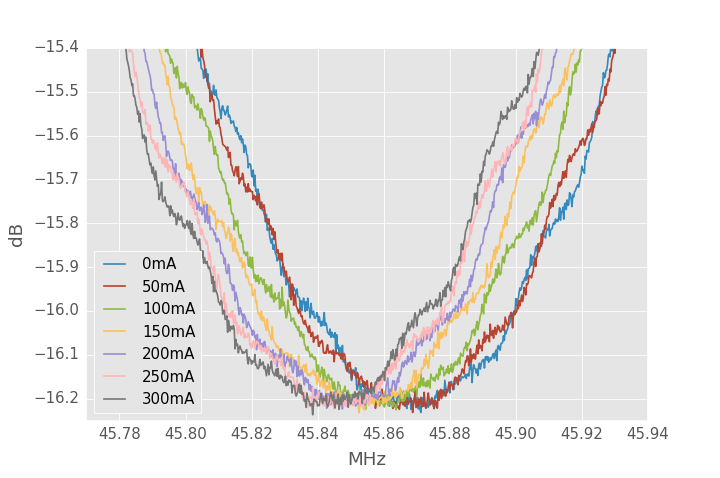}
        \caption{}
        \label{fig:low1}
    \end{subfigure}%
    ~ 
    \begin{subfigure}[t]{0.5\textwidth}
        \centering
        \includegraphics[width=2in]{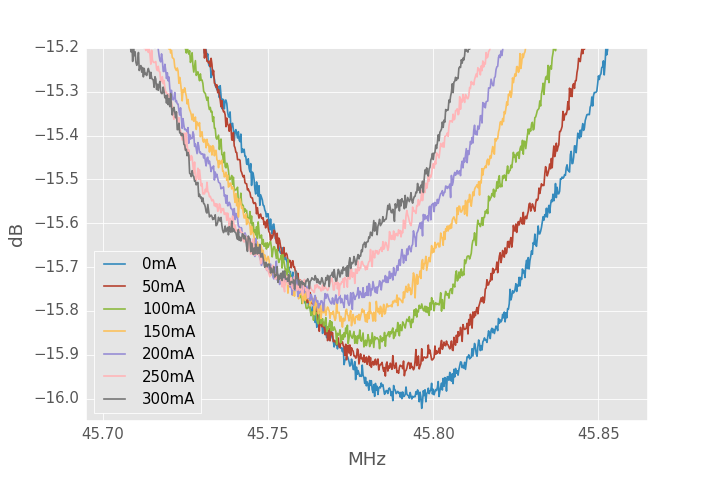}
        \caption{}
        \label{fig:high1}
    \end{subfigure}%
     
    \begin{subfigure}[t]{0.5\textwidth}
        \centering
        \includegraphics[width=2in]{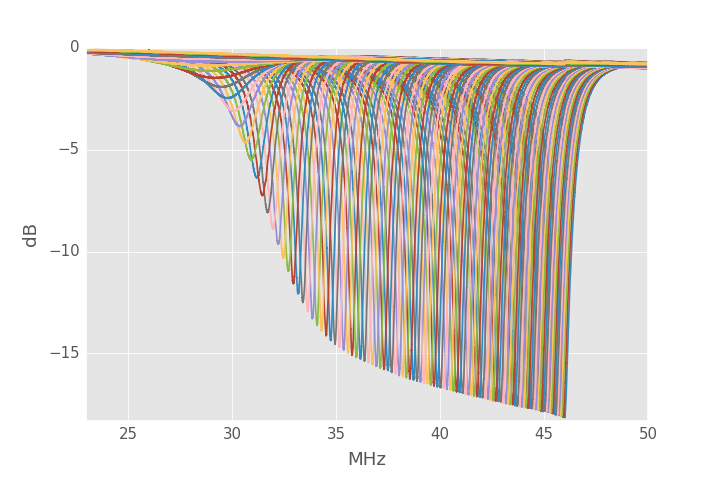}
        \caption{}
        \label{fig:thermal}
    \end{subfigure}%
    ~ 
    \begin{subfigure}[t]{0.5\textwidth}
        \centering
        \includegraphics[width=2in]{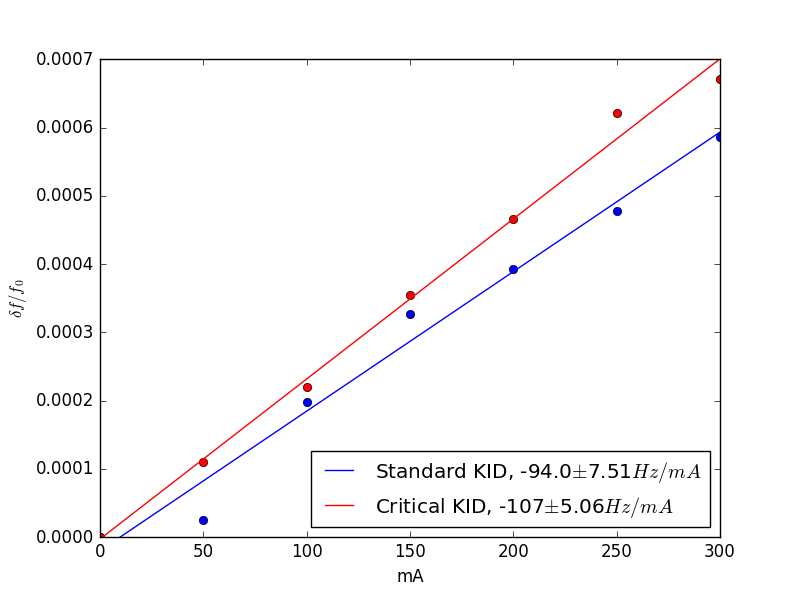}
        \caption{}
        \label{fig:fits}
    \end{subfigure}
    \caption{(a) 
    Resonance shift as a function of LED current for the standard KID mode. 
    (b) 
    Resonance shift as a function of LED current for the critical KID mode. 
    (c) Thermal response of the transmission from the superconducting state at $2.7~K$ to the normal state. (d) Shifts of $94.0\pm7.51\frac{Hz}{mA}$ and $107\pm5.06\frac{Hz}{mA}$ are measured for the standard KID and critical KID mode.
    }
    \label{fig:SNR}
\end{figure*}

When operated in KID mode, the detectors are AC biased with tones at their resonant frequencies of 45.85 and 91.81MHz. We investigate the effect of two regimes of AC biasing. The first regime is a standard, low-current bias of the nanowire such that the critical current is never exceeded. In this regime, the nanowire remains superconducting under optical loading. The incident photons break Cooper pairs and a change in kinetic inductance results in a shift of the resonant frequency as shown in Figure \ref{fig:low1}. A shift of $94\pm7.51\frac{Hz}{mA}$ of LED current is measured. Above a few mA, the LED current is proportional to the emitted power.

Using the measured frequency shift, we can estimate the absorbed optical power by dividing by the responsivity,
\begin{equation}
\frac{df}{dP_{abs}} = \frac{df}{dN_{qp}}\frac{dN_{qp}}{dP_{abs}},
\end{equation}
where $df$ is the frequency shift, $dP_{abs}$ is the absorbed optical power, and $dN_{qp}$ is the number of quasiparticles. 

The shift in frequency due to a change in quasiparticle number density is given by\cite{Phil_Paper,Gao}
\begin{equation}
\frac{df}{dN_{qp}} \cong \frac{-\alpha f_0}{4 N_0\Delta_0}\frac{1}{\Sigma}\left(1+\sqrt{\frac{2\Delta_0}{\pi kT}}\right),
\end{equation}
where $f_0$ is the resonant frequency, $\alpha = 1$ is the kinetic inductance fraction, $N_0$ is the density of states at the Fermi level, $\Delta_0$ is the band-gap energy, $\Sigma$ is the volume of the superconducting nanowire, and $T$ is the base temperature. For $f_0 = 45.87MHz$, $N_0 = 2.55 \times 10^{10}\frac{1}{eV \mu m^3}$ \cite{smirnov}, $\Delta_0 = 1.1$meV, $\Sigma = 0.2866\mu$m$^3$, and $T = 2.7K$, the frequency shift with respect to the generation of a single quasiparticle is $-4\frac{Hz}{qp}$.
The change in quasiparticle number density as a function of absorbed power is given as
\begin{equation}
\frac{dN_{qp}}{dP_{abs}} = \frac{\eta \tau_{\epsilon}}{\Delta_0},
\end{equation}
where $\eta$ is the internal quasiparticle generation efficiency for absorbed optical power and $\tau_{\epsilon}$ is the energy relaxation time. $\tau_{\epsilon}$ is estimated to be $\sim 0.1-0.5$ns based on the time it takes for the energy deposited by the absorbed photon to escape the nanowire \cite{smirnov}. This relaxation time arises from the process of quasiparticle recombination into Cooper pairs and phonons. The phonons have the energy of the superconducting gap such that if they do not thermalize then the net effect should be to effectively increase the energy relaxation time to the phonon escape time. For $\eta \sim 0.5$ and $\Delta_0 = 1.1$meV, the total number of quasiparticles changes by $0.5-2.5 \times 10^{-7}\frac{qp}{eV}$.

The shift in resonant frequency as a function of absorbed power is given as
\begin{equation}
\frac{df}{dP_{abs}} = 2-10 \times 10^{-7} \frac{Hz}{eV~s},
\end{equation}
where since the measured shift is $94\frac{Hz}{mA}$, the absorbed power per $mA$ is $1-5 \times 10^8\frac{eV}{mA~s}$. Each photon at the the operating wavelength of $1.3\mu$m has an energy of $\sim 0.95$eV such that for the standard KID mode we arrive at a photon absorption rate per mA of LED current of
\begin{equation}
\Gamma_{stan} = 1-5 \times 10^8\frac{\rm photons/s}{mA}
\end{equation}
\vspace{-0.5cm}

\subsection{Critical Kinetic Inductance Detector}

The second regime of operation is a high AC-current bias of the nanowire resonator. In this regime, the resonant frequency is not only shifted, but the signal also changes in amplitude as shown in Figure \ref{fig:high1}. This effect is due to the averaging of two states of the resonator. The first state is when the AC bias amplitude is below the critical current and the incident photons act only to change the kinetic inductance and shift the resonant frequency. The second state is a normal state that occurs when a photon acts to drive the nanowire normal as the AC amplitude approaches the critical current.

The fraction of time spent in the normal state can be determined by analyzing the dependence of the change in amplitude of the resonance on the change in current in the LED. For LED currents from $0$ to $50$mA, the resonator amplitude rises from $\sim -16.00$dB to $\sim -15.93$dB. This corresponds to amplitude scale factors of 0.15848 and 0.15977, respectively. Performing the weighted average,
\begin{equation}
0.15977 = \frac{t_{norm}}{T} + 0.15848\Big(1-\frac{t_{norm}}{T}\Big),
\end{equation}
where $t_{norm}$ is the time spent in the normal state and $T$ is the total time of the measurement. The fractional time spent in the normal state is found to be $3 \times 10^{-5}$ per $mA$. 
The pulse recovery time is approximately $200$ns so that the photon count rate where an incident photon causes the resonator to switch to the normal state for the critical KID mode is
\begin{equation}
\Gamma_{crit} = 150\frac{\rm photons/s}{mA}.
\end{equation}
This corresponds to a detection efficiency of approximately $10^{-6}$ of the absorbed photons as estimated from the frequency shift.

\section{Operation in Geiger Mode}

\begin{figure*}[t!]
    \centering
    \begin{subfigure}[t]{0.5\textwidth}
        \centering
        \includegraphics[width=2in]{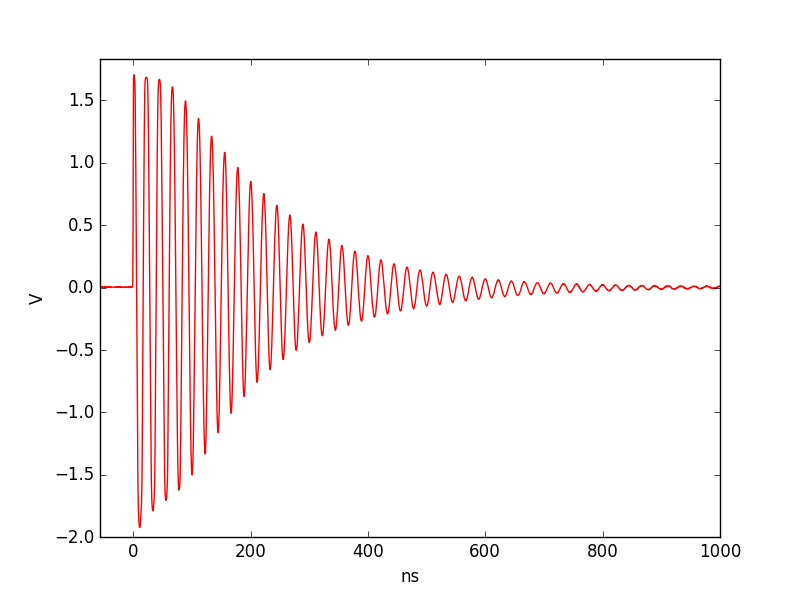}
        \caption{}
        \label{fig:Pulsea}
    \end{subfigure}%
    ~ 
    \begin{subfigure}[t]{0.5\textwidth}
        \centering
        \includegraphics[width=2in]{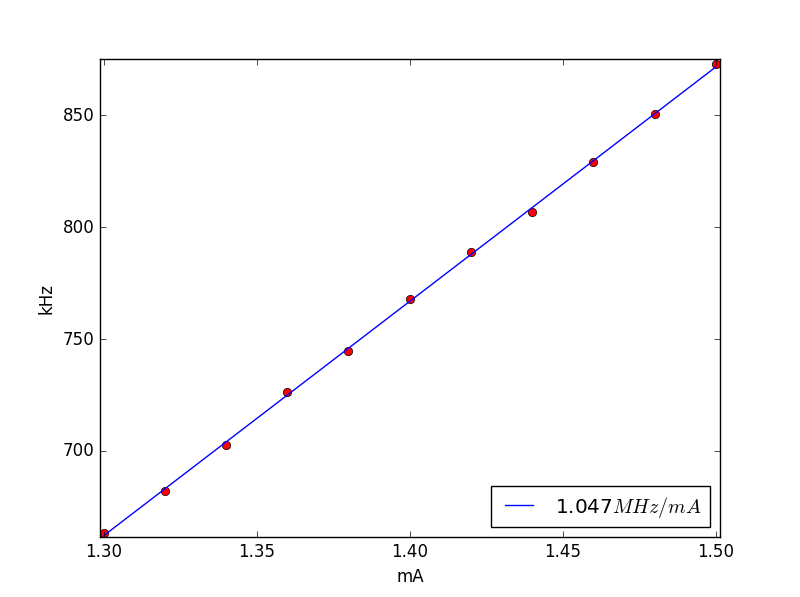}
        \caption{}
        \label{fig:Pulsed}
    \end{subfigure}    
    \caption{(a) Amplified waveform produced by a single-photon incident on the nanowire and traveling through the 10pF coupling capacitor. 
    (b) Count rate as a function of LED current bias.}
    \label{fig:Pulse}
\end{figure*}

When operated as an SPD in Geiger mode, the resonators are DC biased through cryogenic bias tees. At a fixed temperature, the detectors were biased with a DC current and illuminated by an $1.3\mu m$ LED. When a photon is incident on the nanowire, a fast voltage step occurs. Since the nanowire is embedded in a resonant circuit, the signal rings down at the resonant frequency of the detector as shown in Figure \ref{fig:Pulsea}.
The pulses that ring down were rectified with an analog circuit and converted to a standard pulse with an envelop detector
. These pulses are counted with a frequency counter and the resulting count rate as a function of LED power is shown in Figure \ref{fig:Pulsed} for a device DC bias of $5\mu A$. The photon count rate per $mA$ for SPD mode is determined to be 
\begin{equation}
\Gamma_{SPD} = 10^6\frac{\rm photons/s}{mA}
\end{equation}
This corresponds to a detection efficiency of approximately $1\%$ of the absorbed photons.

\section{Discussion}

We have investigated three modes of detection: standard KID, critical KID, and SPD. The frequency shifts for the low-current bias and the high-current bias KID modes were approximately the same. 
The predominant effect of the high-current bias was a decrease in amplitude as the power incident on the device was increased indicating that some of the time the resonator is switching to normal and transmitting the AC signal. 
Compared to DC biasing in SPD mode, AC biasing in critical KID mode has a much lower quantum efficiency due to the small fractional amount of time spent near the critical current of the superconducting nanowire. 

\begin{acknowledgements}
This work is supported in part by NSF AST ATI grant 1509078.
\end{acknowledgements}

\pagebreak

\bibliographystyle{unsrt}
\bibliography{bib}

\begin{thebibliography}{10}

\bibitem{Schroeder2016}
Edward Schroeder, Philip Mauskopf, Genady Pilyavsky, Adrian Sinclair, Nathan
  Smith, Sean Bryan, Hamdi Mani, Dmitry Morozov, Karl Berggren, Di~Zhu,
  Konstantin Smirnov, and Yuriy Vakhtomin.
\newblock On the measurement of intensity correlations from laboratory and
  astronomical sources with spads and snspds.
\newblock volume 9907 of {\em Proc. SPIE}, pages 99070P--99070P--13, 2016.

\bibitem{2012SuScT..25f3001N}
C.~M. {Natarajan}, M.~G. {Tanner}, and R.~H. {Hadfield}.
\newblock {Superconducting nanowire single-photon detectors: physics and
  applications}.
\newblock {\em Supercond. Sci. Technol.}, 25(6):063001, June 2012.

\bibitem{Zhu:16}
Di~Zhu, Hyeongrak Choi, Tsung-Ju Lu, Qingyuan Zhao, Andrew Dane, Faraz Najafi,
  Dirk~R. Englund, and Karl Berggren.
\newblock Superconducting nanowire single-photon detector on aluminum nitride.
\newblock In {\em Conference on Lasers and Electro-Optics}, page FTu4C.1.
  Optical Society of America, 2016.

\bibitem{0953-2048-28-11-114003}
A~Engel, J~J Renema, K~Il’in, and A~Semenov.
\newblock Detection mechanism of superconducting nanowire single-photon
  detectors.
\newblock {\em Supercond. Sci. Technol.}, 28(11):114003, 2015.

\bibitem{0957-4484-21-44-445202}
Anthony~J Annunziata, Daniel~F Santavicca, Luigi Frunzio, Gianluigi Catelani,
  Michael~J Rooks, Aviad Frydman, and Daniel~E Prober.
\newblock Tunable superconducting nanoinductors.
\newblock {\em Nanotechnology}, 21(44):445202, 2010.

\bibitem{7399730}
S.~Doerner, A.~Kuzmin, S.~Wuensch, K.~Ilin, and M.~Siegel.
\newblock Operation of superconducting nanowire single-photon detectors
  embedded in lumped-element resonant circuits.
\newblock {\em IEEE Trans. Appl. Supercond.}, 26(3):1--5, April 2016.

\bibitem{doi:10.1063/1.4993779}
S.~Doerner, A.~Kuzmin, S.~Wuensch, I.~Charaev, F.~Boes, T.~Zwick, and
  M.~Siegel.
\newblock Frequency-multiplexed bias and readout of a 16-pixel superconducting
  nanowire single-photon detector array.
\newblock {\em Appl. Phys. Lett.}, 111(3):032603, 2017.

\bibitem{6459543}
M.~Hofherr, M.~Arndt, K.~Il'in, D.~Henrich, M.~Siegel, J.~Toussaint, T.~May,
  and H.~G. Meyer.
\newblock Time-tagged multiplexing of serially biased superconducting nanowire
  single-photon detectors.
\newblock {\em IEEE Trans. Appl. Supercond.}, 23(3):2501205--2501205, June
  2013.

\bibitem{Zhao2017}
Qing-Yuan Zhao, Di~Zhu, Niccol{\`o} Calandri, Andrew~E. Dane, Adam~N.
  McCaughan, Francesco Bellei, Hao-Zhu Wang, Daniel~F. Santavicca, and Karl~K.
  Berggren.
\newblock Single-photon imager based on a superconducting nanowire delay line.
\newblock {\em Nat. Photon.}, 11(4):247--251, Apr 2017.
\newblock Article.

\bibitem{Phil_Paper}
P.D. {Mauskopf}.
\newblock {Transition Edge Sensors and Kinetic Inductance Detectors in
  Astronomical Instruments}.
\newblock {\em to appear in PASP}, 2017.

\bibitem{Gao}
J.~Gao, M.~R. Vissers, M.~O. Sandberg, F.~C.~S. da~Silva, S.~W. Nam, D.~P.
  Pappas, D.~S. Wisbey, E.~C. Langman, S.~R. Meeker, B.~A. Mazin, H.~G. Leduc,
  J.~Zmuidzinas, and K.~D. Irwin.
\newblock A titanium-nitride near-infrared kinetic inductance photon-counting
  detector and its anomalous electrodynamics.
\newblock {\em Applied Physics Letters}, 101(14):142602, 2012.

\bibitem{smirnov}
K.~V. Smirnov, A.~V. Divochiy, Yu.~B. Vakhtomin, M.~V. Sidorova, U.~V. Karpova,
  P.~V. Morozov, V.~A. Seleznev, A.~N. Zotova, and D.~Yu. Vodolazov.
\newblock Rise time of voltage pulses in nbn superconducting single photon
  detectors.
\newblock {\em Applied Physics Letters}, 109(5):052601, 2016.

\end{thebibliography}

\end{document}